\documentclass[aps,twocolumn,prb,amsmath,amssymb,groupedaddress,floatfix]{revtex4}

\usepackage{graphicx}
\usepackage{dcolumn}
\begin{document}

\bibliographystyle{revtex}


\title{Theory of helical spin crystals: phases, textures and properties}


\author{B. Binz}
\email{binzb@berkeley.edu}
\affiliation{Department of Physics, University of California, 366 Le Conte \# 7300, Berkeley, CA 94720 - 7300, USA}
\author{A. Vishwanath}
\affiliation{Department of Physics, University of California, 366 Le Conte \# 7300, Berkeley, CA 94720 - 7300, USA}


\date{\today}

\newcommand{\eref}[1]{(\ref{#1})}

\newcommand{\bv}[1]{{\bf #1}}
\newcommand{\uv}[1]{{\bf \hat #1}}

\renewcommand{\Re}{\mathrm Re}
\newcommand{\be}{\begin{equation}}
\newcommand{\ee}{\end{equation}}

\begin{abstract}
Motivated by recent experiments on the itinerant helimagnet MnSi, we study
general properties of helical spin crystals - magnetic structures obtained
by superposing distinct spin spirals. 
An effective Landau description of helical spin crystals
is introduced and simple rules for stabilizing various spin crystal
structures over single spirals are established. Curious properties of the
magnetic structures so obtained, such as symmetry stabilized topological
textures and missing Bragg reflections are pointed out. The
response of helical spin crystals to crystalline anisotropy, magnetic
field and non-magnetic disorder are studied, with special reference to the
bcc1 spin structure, a promising starting point for discussing the  'partial order'
phases seen at high pressure in MnSi. Similar approaches may be applied to other crystallization problems
such as Larkin-Ovchinnikov-Fulde-Ferrel states in spin-imbalanced superconductors.
\end{abstract}

\maketitle 

\section{Introduction}

Long period helical spin order resulting from
Dzyaloshinskii-Moriya (DM) spin-orbit
coupling\cite{dzyaloshinskii58} in non-centrosymmetric
itinerant magnets (e.g. MnSi, Fe$_x$Co$_{1-x}$Si, FeGe with crystal
structure B20) has been intensively studied in the
past.\cite{dzyaloshinskii64,ishikawa76, moriya76,nakanishi80,bak80,beille83,lebech89,ishimoto95,uchida06}
One material of this family, MnSi, has attracted recent interest
because of its puzzling behavior under applied hydrostatic
pressure.\cite{pfleiderer97,thessieu97} 
 A state with non-Fermi liquid transport
properties\cite{pfleiderer01} is obtained over a wide range of
pressures above a critical threshold $p_c$. 
Beginning at the same critical pressure, but over a smaller
pressure range,  magnetic 'partial order' is observed in neutron
scattering experiments.\cite{yu04,pfleiderer04} While usual helical
order gives rise to sharp Bragg peaks in neutron scattering
(corresponding to the periodicity of the helical spin-density wave),
'partial order' is characterized by a neutron scattering signal which
is smeared over a wavevector sphere rather than localized at
discrete points in reciprocal space.

Recent theoretical work on electronic properties, critical fluctuations and
collective modes of helimagnets 
are in Refs.~\cite{fischer04,grigoriev05,belitz06}. Theoretical
proposals for the high pressure state of MnSi have invoked proximity to a
quantum multi-critical point \cite{schmalian04} or magnetic liquid-gas transitions.\cite{tewari05} Closest in spirit to our approach are the skyrmion-like magnetic patterns studied in Refs.~\cite{bogdanov05,fischer06}.

Recently, we have proposed a novel kind of magnetic order, the
helical spin crystal, as a promising starting point for a theory of
'partial order'.\cite{binz06} Helical spin crystals are magnetic
patterns, which are obtained by superposition of several helical
spin-density waves which propagate in different directions. There is a substabtial resemblance to multi-$k$ magnetic structures (also known as multiple-$q$ or multiple spin density wave states).\cite{mukamel76,jo83,forgan89,longfield02,steward04} But in contrast to most other magnetic multi-$k$  systems,  in helical spin crystals the ordering wavevectors are selected  from an infinite number of degenerate modes lying on a sphere in reciprocal space - a process analogous to the crystallization of liquids.

In this
work, we present a detailed theory of such structures. The
stability, structure and distinctive properties of such states are
described, and the consequences of 
 coupling to non-magnetic disorder is discussed.

The paper is organized as follows. First, we review the standard
theory of helimagnetism in Section \ref{GLsection}, finishing with a
short remark about more general helical magnetic states.
 Then, the theory of helical spin crystals is developed.
The requirements
to energetically stabilize  helical spin crystal states are
investigated in Section \ref{energetics}. The analysis works in two directions. First, we establish a phase diagram in terms of natural parameters which tune the interaction between helical modes and second, we give simple rules to construct  model interactions which stabilize a large class of helical spin crystals. The remaining parts of the
paper are dedicated to extracting testable consequences of these
novel magnetic states. In Section \ref{structure},  we give a
description of the most prominent spin crystals which emerge from
our energetic analysis in terms of their symmetry. It is shown that the symmetry  of the magnetic state may stabilize topological textures like merons and anti-vortices which are otherwise not expected to be stable in the present context, given the order parameter and dimensionality of the system.
Symmetry also determines, which higher-harmonics Bragg peaks these structures would produce.
We subsequently study the response of helical spin crystals with respect to
different perturbations in Section \ref{response}. For example, sub-leading spin orbit coupling (crystal anisotropy) locks the magnetic crystal to the
 underlying atomic lattice and thus determines the location of magnetic Bragg peaks. 
We also study the response to an
external magnetic field which, apart from producing a uniform
magnetic moment, also leads to distinctive distortions of the helical
magnetic structure, which could be observable by neutron scattering. 
Finally, in Section \ref{disorder} we investigate the implications of
non-magnetic impurities, which are expected to destroy long-range
magnetic order and produce diffuse scattering.

\section{Landau-Ginzburg theory of helimagnetism}\label{GLsection}

For a cubic magnet without a center of inversion,  the Landau-Ginzburg free energy to quadratic order in the magnetization $\bv M(\bv r)$ is
\be
F_2=\left\langle r_0 \bv M^2 + J (\partial_\alpha M_\beta)(\partial_\alpha M_\beta)+2D\bv M\cdot(\bv \nabla\times\bv M)\right\rangle,\label{F2}
\ee
where $\left\langle\ldots\right\rangle$ indicates sample averaging,  $r_0,J,D$ are parameters ($J>0$)  and Einstein summation is understood. The last term of Eq.~\eref{F2} is the DM interaction, which is odd under spatial inversion and originates in spin-orbit coupling.\cite{dzyaloshinskii58} Fourier transformation, $\bv M(\bv r)=\sum_{\bv q} \bv m_{\bv q}e^{i\bv q\cdot\bv r}$ with $\bv m_{-\bv q}=\bv m^*_{\bv q}$, leads to
\be
F_2=\sum _{\bv q} \left[\left(r_0+Jq^2\right)|\bv m_{\bv q}|^2+2D \bv m_{\bv q}^*\cdot\left(i\bv q\times\bv m_{\bv q}\right)\right].
\ee
Clearly, the energy is minimal for circularly polarized spiral modes,
 where  $\nabla\times\bv M$ points in the direction of $-D\,\bv M$. For such modes,
\be
F_2=\sum _{\bv q} r(q)|\bv m_{\bv q}|^2,
\ee
where $r(q)=r_0-JQ^2+J(q-Q)^2$ with $Q=|D|/J$.
The Gaussian theory thus determines both the chirality of low-energy helical modes and their wavelength $\lambda=2\pi/Q$. The latter  is typically long (between $180\mathring{A}$ in MnSi and $2300\mathring{A}$ in Fe$_{0.3}$Co$_{0.7}$Si), reflecting the smallness of spin-orbit coupling effects compared to exchange.
 However, no preferred spiraling direction is selected by Eq.~\eref{F2}, since $F_2$ is rotation-invariant. Cubic anisotropy terms which break this invariance are of higher order in the spin-orbit interaction and therefore small. We neglect them for the moment and reintroduce them later.

The isotropic Gaussian theory leaves us with an infinite number of modes which become soft as $r(Q)\to0$. They consist of helical spin-density waves with given chirality (determined by the sign of $D$), whose wave-vectors lie on a sphere $|\bv q|=Q$ in reciprocal space.
 Each of these helical modes is determined by an amplitude and a phase.  Hence, for each point  $\bv q$ on the sphere, we define a complex order parameter $\psi_{\bv q}$ (with $\psi_{-\bv q}=\psi^*_{\bv q}$) through
\be
\bv m_{\bv q}=\frac12\psi_{\bv q}(\uv\epsilon'_{\bv q}+i \uv\epsilon''_{\bv q}),\label{psis}
\ee
where $\uv \epsilon'_{\bv q}$, $\uv \epsilon''_{\bv q}$, and $\uv q$, are mutually orthogonal unit vectors (with a defined handedness, given by the sign of $D$). Obviously, changing the phase of $\psi_{\bv q}$ is equivalent to rotating $\uv \epsilon'_{\bv q}$ and $\uv \epsilon''_{\bv q}$ around $\uv q$. The phase of $\psi_{\bv q}$  is thus only defined relative to some initial choice of $\epsilon'_{\bv q}$. The neutron scattering intensity is proportional to $|\uv q\times \bv m_{\bv q}|^2=1/2|\psi_{\bv q}|^2$, independent of the phase. Changing the phase of $\psi_{\bv q}$ is also equivalent to translating $\bv M(\bv r)$ along  $\uv q$.

In the following, we  study minima of the free energy in the ordered phase [$r(Q)<0$].
These depend on the interactions between  degenerate modes (i.e., free energy contributions, which are quartic or higher order in $\bv M$).  We only consider interactions which, as $F_2$, have full rotation symmetry and we will include the weak crystal anisotropy last. The most general quartic term which has full rotation symmetry (transforming space and spin together) is of the form
\be
 F_4=\!\!\sum_{\bv q_1,\bv q_2,\bv q_3}\!\!U(\bv
q_1,\bv q_2,\bv q_3)\left(\bv m_{\bv q_1}\cdot\bv m_{\bv
q_2}\right)\left(\bv m_{\bv q_3}\cdot\bv m_{\bv
q_4}\right),\label{F4}
\ee
 with $\bv q_4=-(\bv q_1+\bv q_2+\bv q_3)$.

\subsection{Single-spiral state}

For example, if $U(\bv q_1,\bv q_2,\bv q_3)$ is a  constant, then $F_4\propto \left\langle \bv M^4\right\rangle$.
If the interaction depends only on the local magnetization amplitude, i.e., in general if $F=F_2+\left\langle f(\bv M^2)\right\rangle$ for some function $f$, then the absolute minimum of $F$ is given by a single-spiral state (also known as helical spin density wave) $\bv M(\bv r)=\bv m_{\bv k}e^{i\bv k\cdot\bv r}+\bv m^*_{\bv k}e^{-i\bv k\cdot\bv r}$, where a single pair of opposite momenta $\pm \bv k$ is selected. 
 To proof this, we write $F$ as
\be
\sum_{\bv q} [r(q)-r(Q)]\,|\bv m_{\bv q}|^2 + \left\langle r(Q)\bv M^2+f(\bv M^2)\right\rangle.\label{proof}
\ee
In the single-spiral state, $\bv M^2$ is constant in space and it minimizes the first and the second term of Eq.~\eref{proof} independently. Therefore, no other magnetic state can be lower in energy.

Because $Q$ is small,  the relevant wavevectors entering Eq.~\eref{F4} are also small and  $U(\bv q_1,\bv q_2,\bv q_3)$ is effectively
 close to a constant. Therefore, the single-spiral state, as observed in Fe$_x$Co$_{1-x}$Si, FeGe and in MnSi at ambient pressure, is the most natural helical magnetic order from the point of view of Landau theory.

\subsection{Linear superpositions of single-spiral states}\label{continuous}

Motivated by the phenomenology of  'partial order', we will now  extend the theory beyond this standard solution. We speculate that $U(\bv q_1,\bv q_2,\bv q_3)$ is not constant, such that $F_4$  favors a  linear superpositions of multiple  spin-spirals with different wave-vectors on the sphere of degenerate modes $|\bv q|=Q$.

One may first speculate about magnetic patterns whose Fourier transform
is non-zero everywhere on the wave-vector sphere and
 peaked infinitely sharply perpendicular to the sphere, i.e.
\be
|\psi_{\bv q}|^2\propto \delta(|\bv q|-Q)\label{po-literally}
\ee
[see Eq.~\eref{psis}]. This idea turns out to be complicated for at least two reasons.

The first complication is that there is no continuous way of attributing a finite-amplitude spiral mode to each point on the wavevector sphere. This can be seen by noting that $\uv \epsilon'_{\bv q}$ [Eq.~\eref{psis}] is a tangent vector field on the sphere. Thus, it cannot be continuous (impossibility of combing a hedgehog).\cite{mermin79} Thus there is no ``uniform''  superposition of helical modes on the sphere. The problem of singularities can be avoided if one assumes a $\psi_{\bv q}$ with point nodes on the sphere.

The second complication is that   higher harmonics would result in a broadening of the delta-function in Eq.~\eref{po-literally}. This is seen as follows. Consider three momenta $\bv q_1, \bv q_2, \bv q_3$ on the wavevector sphere and $\bv q_4$ which is off the sphere. The non-vanishing modes $\bv m_{\bv q_1},\bv m_{\bv q_2}, \bv m_{\bv q_3}$  couple linearly to  $\bv m_{\bv q_4}$ via Eq.~\eref{F4} and thus induce a higher harmonic ``off-shell'' mode $\bv m_{\bv q_4}\neq0$. Since this happens for every point away from the sphere,
 the effect is an intrinsic broadening of the peak in $|\psi_{\bv q}|^2$, in contradiction with the initial  assumption of Eq.~\eref{po-literally}.

\section{Energetics of helical spin crystals}\label{energetics}

In the following, we study magnetic structures which are superpositions of a {\em  finite} number of degenerate helical modes $\psi_j$ with wavevectors $\pm\bv k_j$, $j=1,\ldots,N$. We call the resulting states helical spin crystals, because of the analogy with weak crystallization theory of the solid-liquid transition.\cite{brazovskii87}

\subsection{Structure of the quartic interaction}\label{section_int}

\begin{figure}
\includegraphics[scale=0.6]{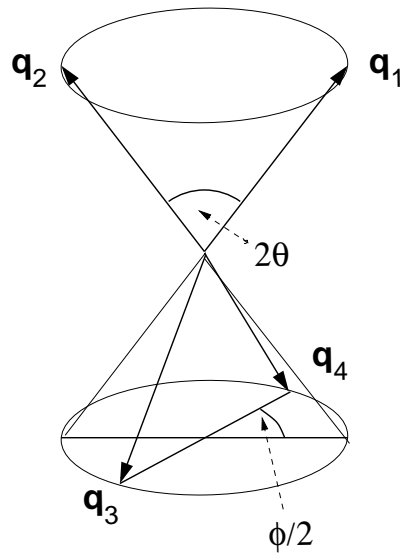}
\caption{The set of quartets $(\bv q_1,\ldots,\bv q_4)$  satisfying  $|\bv q_i|=Q$ and  $\bv q_1+\ldots+\bv q_4=0$, modulo global rotations, may be parameterized by two angles $\theta$ and $\phi$ as shown in this figure.  \label{angles}}
\end{figure}

We assume that $F_4$ is small, and that its main effect is to provide an
interaction between the modes which are degenerate under $F_2$.
Thus, the relevant terms of $F_4$ are those with $|\bv q_1|= |\bv q_2|=|\bv
q_3|=|\bv q_4|=Q$. This phase-space constraint and  rotational symmetry implies that the coupling function $U$ depends only on two relative angles between the momenta
\be
\left.U(\bv q_1,\bv q_2,\bv q_3)\right|_{|\bv q_i|=Q}=U(\theta,\phi),\label{effint}
\ee
 where we have chosen the following parameterization:
\be
\begin{split}
2\,\theta&=\arccos(\uv q_1\cdot\uv q_2)\\
\phi/2&=\arccos\left[\frac{(\uv q_2-\uv q_1)\cdot \uv q_3}{1-\uv q_1\cdot\uv q_2}\right].
\end{split}\label{thph}
\ee
 Geometrically,
$\phi/2$ is the angle between the two planes
spanned by $(\bv q_1, \bv q_2)$ and  $(\bv q_3,\bv q_4)$ (Fig.~\ref{angles}). In the special case $\bv q_1+\bv q_2=0$, it
 becomes the angle between $\bv q_2$ and  $\bv q_3$. This mapping allows $\theta$ and $\phi$ to be interpreted as the polar and azimuthal angles of a sphere and the coupling $U(\theta,\phi)$ is a function on that sphere.   Since it describes an effective coupling between modes on the wavevector sphere, the coupling  $U(\theta,\phi)$ has a status similar to that of Fermi liquid parameters in the theory of metals.

The Landau free energy $F=F_2+F_4$ of helical spin crystal states is calculated as follows. Obviously,
\be
F_2=r(Q)\sum_j|\psi_j|^2.\label{F2psi}
\ee
The quartic term, Eq.~\eref{F4}, may be split into three distinct contributions $F_4=F_{s}+F_{p}+F_{\rm nt}$. The first term, $F_{s}$, is the self-interaction of each spiral mode with itself:
\be
F_{s}=U_{s}\sum_j|\psi_j|^4,\label{Fs}
\ee
where $U_{s}=U\!(\theta\!=\!\pi/2,\phi\!=\!0)$. A minimum requirement for stability of the  theory is $U_{s}>0$.

The next term, $F_{p}$, stems from pairwise interactions between modes. It is of the form
\be
F_{p}=2\sum_{i<j} V_{p}(\theta_{ij})\, |\psi_i|^2|\psi_j|^2,\label{Fp}
\ee
where  $2\theta_{ij}=\arccos(\uv k_i\cdot\uv k_j)$ and
\be
V_{p}(\theta)=U\!(\frac\pi2,4\theta)+\sin^4\!\theta\,U\!(\theta,0)+\cos^4\!\theta\,U\!(\frac\pi2\!-\!\theta,0).\label{V}
\ee
Since $\lim_{\theta\to0}V_{p}(\theta)=2U_{s}$ is large and positive, $F_{p}$ provides an efficient repulsion of modes which are too close on the wavevector sphere. Energetically stable superpositions of spirals require substantially smaller values of
 $V_{p}(\theta)$ and therefore big-enough angles between them (Section \ref{exact}). This ``mode repulsion'' suggests that a continuous distribution of Fourier modes on the wavevector sphere, as discussed in the  Section \ref{continuous}, is difficult to stabilize energetically by  slowly varying coupling functions $U(\theta,\phi)$. However, singular behavior of the coupling  function is conceivable when discussing effective interactions of low energy modes.\cite{binz02}

Finally, the non-trivial quartic term $F_{\rm nt}$ stems from  quartets of modes, whose wavevectors sum up to zero. The geometry of four equal-length  wavevectors summing to zero is depicted in Fig.~\ref{angles}.  A  quartet-contribution occurs for $0<2\theta<\pi$ and $0<\phi/2<\pi$. In this case, the eight wavevectors $\pm\bv k_j$  form the vertices of a cuboid. Hence,
\be
F_{\rm nt}={\sum_{j_1<\ldots<j_4}}^{\!\!\!\!\!\!\prime}\,\, F_{j_1,j_2,j_3,j_4},\label{Fq}
\ee
where the summation is over such quartets. For example, if $\bv k_{j_1}+\bv k_{j_2}+\bv k_{j_3}+\bv k_{j_4}=0$, there is a term $ F_{j_1,j_2,j_3,j_4}\propto\psi_{j_1}\psi_{j_2}\psi_{j_3}\psi_{j_4}$. 
 The algebra of these terms, which depend on the relative phases of  modes, is rather lengthy and obviously  depending on the phase convention used to define the $\psi$-variables.

The term $F_{\rm nt}$ fixes the relative phases of $\psi_j$  to minimize the energy.  
 If there is no frustration between minimizing each individual quartet-term, the phases arrange in such a way that all $F_{j_1,j_2,j_3,j_4}\leq 0$. It follows that $F_{\rm nt}\leq0$, after minimization for the $\psi$-phases.  A special case is the quartet of modes, whose wavevectors $\pm k_j$ form the vertices of a cube (e.g. modes along  $\langle111\rangle$). 
  In this case,  $F_{j_1,j_2,j_3,j_4}=0$ independently of the coupling function $U(\theta,\phi)$ as a consequence of rotational symmetries.\cite{ashvin06} This is the reason why no such quartet term appears in the theory of Bak and Jensen.\cite{bak80} As a consequence, a superposition of four modes along all $\langle111\rangle$ directions is necessarily unstable towards  shifting the wavevectors away from the perfect cubic configuration in order to gain energy from $F_{\rm nt}$ (see description of fcc$^*$  below).

\subsection{Phase diagram}

If $U(\theta,\phi)$ is only slowly varying, it is justified to expand it in spherical harmonics ($Y_{lm}$). That is,
\be
\begin{split}
U(\theta,\phi)&=U_0+U_{11}\sin\theta\cos\phi\\
& \quad +U_{20}(3\cos^2\theta-1)+U_{22}\sin^2\theta\cos2\phi,
\end{split}\label{Ulm}
\ee
where we retained all terms with $l\leq2$, which satisfy the relation $U(\theta,\phi)=U(\pi-\theta,\phi)=U(\theta,2\pi-\phi)$.

Different quartic expressions in terms of the real-space magnetization $\bv M(\bv r)$ may lead to the same projected coupling $U(\theta,\phi)$. For example, the  three  terms
\be
\left\langle W\,\bv M^4+W'\,\left[\bv\nabla\left(\bv M^2\right)\right]^2+W''\, (\partial_\alpha M_\beta)(\partial_\alpha M_\beta)\bv M^2\right\rangle \label{W'}
\ee
lead to $U(\bv q_1,\bv q_2,\bv q_3)=W+W'(\bv q_1+\bv q_2)^2-W''\bv q_1\cdot\bv q_2$. When projected onto the wavevector sphere, they generate only two spherical harmonics, namely
\be
\begin{split}
U_0&=W+\frac13(4W'+W'')Q^2,\\
U_{20}&=\frac23(2W'-W'')Q^2.
\end{split}\label{dict}
\ee
The terms $W$ and $W'$ have  intuitive interpretations: $W$ restricts  the magnetization amplitude,  whereas $W'$ favors or disfavors modulations of $\bv M^2$, depending on the sign.\footnote{Whenever $W'<0$, Eq.~\eref{W'} becomes negative for a rapidly varying $\bv M^2$ and the quartic theory is thus unstable. We consciously ignore this instability, since we only study local minima at small $\bv M^2$, where $F_2$ is dominant and forbids rapid variations.} In the following, we set $W''=0$ and use the four parameters $W,W',U_{11}$ and $U_{22}$ to tune the interaction $U(\theta,\phi)$.

\begin{figure}
\includegraphics[scale=0.6]{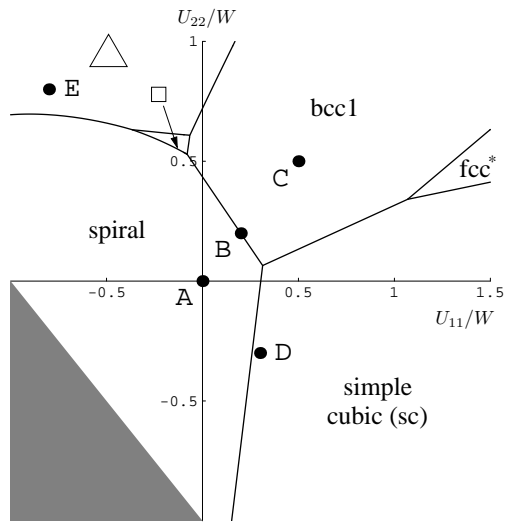}
\caption{Phase diagram for  $W>0$ and $W'=0$.  In the grey region, $F4<0$ and the quartic theory is unstable. The various phases are explained in the text. In contrast to our earlier use of these symbols,\cite{binz06}  ``$\bigtriangleup$'' and ``$\square$''  denote general states with 3 and 4 helical modes, respectively.
\label{pdiag0}}
\end{figure}

\begin{figure}
\includegraphics[scale=0.6]{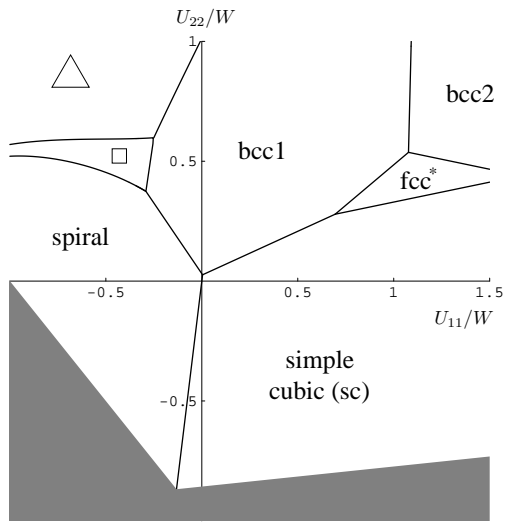}
\caption{Same as Fig.~\ref{pdiag-2} with $Q^2W'=-0.5W$. The point $U_{11}=U_{22}=0$ is now at the phase boundary between spiral order and simple cubic.
\label{pdiag-2}}
\end{figure}

It is hard to find the exact global minimum of $F_2+F_4$ for a general coupling function $U(\theta,\phi)$. We therefore restrict ourselves to a certain variational  class of magnetic states and determine the minimum within this class. Recently,\cite{binz06} we studied  only states which can be obtained by superposition of those six modes which  propagate along the  $\langle110\rangle$  directions (6-mode model).
Here, we explore a much broader class of states. We include
\begin{enumerate}
\item any superposition of those 13 spin-spirals which propagate along the directions\footnote{Only the relative orientation between these modes enters the calculation, not their orientation with respect to the atomic lattice, since crystal anisotropy is neglected.} $\langle111\rangle$,  $\langle110\rangle$ and  $\langle100\rangle$ (13 complex variables $\psi_j$ as variational parameters),
\item  any superposition of up to four spirals with arbitrary propagation direction (four  variables $\psi_j$ and five independent angles between wavevectors as  variational parameters).
\end{enumerate}
Within these constraints, we  have computed\footnote{To obtain the phase diagram of Figs.~\ref{pdiag0} and \ref{pdiag-2}, we have used the approximation $(1+x)^{5/2}+(1-x)^{5/2}\approx 2+(4\sqrt{2}-2)x^2$ for $x=\cos\theta$ in the evaluation of $V_{p}(\theta)$. The error is smaller than  0.7\%.} a phase diagram as a function of the coupling parameters $W,W',U_{11}$ and $U_{22}$, shown in Figs.~\ref{pdiag-2} and \ref{pdiag0}.

\begin{figure}
\includegraphics[scale=0.6]{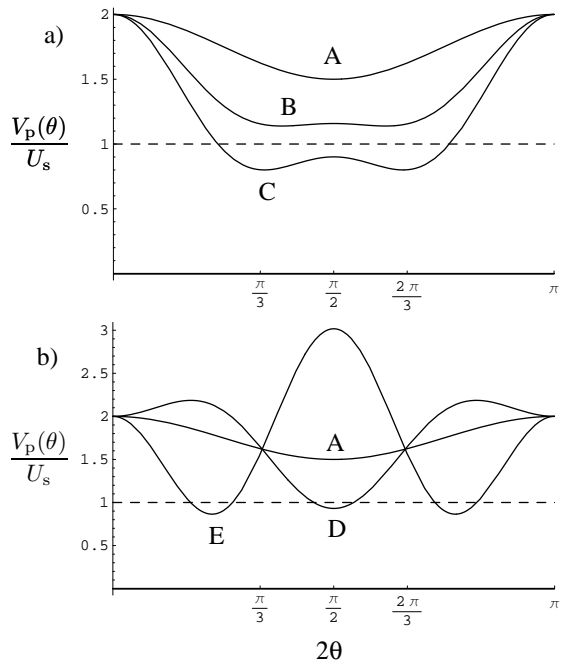}
\caption{a) and b) For the locations $A,\ldots,E$ in the phase diagram of Fig.~\ref{pdiag0},
the ratio between the pair interaction $V_{p}(\theta)$ and the self-interaction $U_{s}$ is plotted as a function of  the angle $2\theta$ (the angle between propagation directions of modes). A: single-spiral state; B: phase boundary between single-spiral and bcc1; C: bcc1; D: sc; E: $\bigtriangleup$.
\label{Vpfig}}
\end{figure}

All magnetic ground states are equal-amplitude superpositions of 1,2,3,4 or 6 spiral modes.
 The body-centered cubic (bcc) states are superpositions of all  $\langle110\rangle$-modes.
 bcc1 and bcc2 differ by the relative phases of the six interfering helical modes (see Section \ref{bcc1}).  The simple cubic (sc) crystal consists of three mutually orthogonal spirals (e.g. along all $\langle100\rangle$ directions). A face-centered cubic (fcc) helical spin crystal is obtained by superposing all four  $\langle111\rangle$-modes. However, the  ground state is not fcc but a small distortion of it: fcc$^*$. In fcc$^*$, the wavevectors are  shifted slightly  away from $\langle111\rangle$  in order to gain energy from the quartet-term $F_{\rm nt}$.\footnote{The shift
 to directions $[\pm\sqrt{1+\delta}\,\pm\!\!\sqrt{1+\delta}\,\sqrt{1-2\delta}\,]$  is very small with $\delta\sim 0.02$ and even without the deformation, fcc would still beat the neighboring phases in the fcc$^*$-region.}  The symbols ``$\bigtriangleup$'' and ``$\square$'' are used differently here than in our previous paper.\cite{binz06} Here, ``$\square$''  stands for a superposition of four modes with wavevectors as shown in Fig.~\ref{angles} with  $\phi/2=\pi/2$. Hence,  wavevectors $\pm k_j$ form a square cuboid and allow for  one quartet-term $F_{\rm nt}$. The angle $2\theta$ changes as a function of interaction parameters within the range  $0.24\pi<2\theta<0.38\pi$.
Finally, the phase ``$\bigtriangleup$'' consists of  three modes. The wavevectors $\bv k_1,\bv k_2,\bv k_3$ point to the vertices of an equilateral triangle on the sphere, whose size is determined by the requirement that the mutual angle  between two wavevectors, $2\theta$, minimizes Eq.~\eref{V}. The angle $2\theta$ is parameter-dependent  and lies in the  range $0.14\pi<2\theta<0.24\pi$.

As expected, a negative $W'$ [Eq.~\eref{W'}] favors multi-mode spin crystal states with varying magnetization amplitude relative to the spiral state with constant $\bv M^2$ (compare Figs.~\ref{pdiag0} and \ref{pdiag-2}). Positive $W'$ has the opposite effect, and enhances the region of the spiral phase (not shown). The term $W'$ alone (i.e., with $U_{11}=U_{22}=0$) stabilizes sc in the regime $Q^2W'<-W/2<0$. However, a small positive $U_{22}$ is sufficient to favors bcc1 over sc.

In conclusion, we observe that  two helical spin crystals, bcc1 and sc, appear adjacent to the single-spiral state and are stable at relatively small values of  $Q^2W'$, $U_{11}$ and $U_{22}$.
 In the following, we study the properties of the bcc1, and sc states since they are the most likely candidates of helical spin crystals from the point of view of energetics.

\subsection{Model interactions with exact ground states}\label{exact}

In the preceding Section, we established a variational phase diagram for ``natural'', i.e., slowly varying coupling functions $U(\theta,\phi)$. Most phases in this phase diagram, (``spiral'', sc, bcc1, bcc2, fcc and ``$\bigtriangleup$'') can  be shown to be the {\em exact} global minima for some fine-tuned model interaction, that are constructed below.

Let us consider the toy model $F_{{\rm toy}}=F_2+F_{s}+F_{p}$ [Eqs.~(\ref{F2psi}-\ref{Fp})] where the quartet term $F_{\rm nt}$ is dropped and we replace $V_{p}(\theta)$ by a constant $V$. In this model,  local minima with $N$ non-vanishing modes ($N\geq1$) must have equal amplitudes  $|\psi_j|^2=|r|/(U_{s}-V+N V)/2$ and the minimum energy with $N$ modes is
\be
F_{{\rm toy},N}=-\frac14\,\frac{r^2}{V+\frac{U_{s}-V}N}.
\ee
There are three regimes. If $0<U_{s}<V$, the ground state is the single-mode phase with $N=1$ (all but one $\psi$ are zero). For $0<V<U_{s}$, the energy is monotonically decreasing with $N$. It means that the system includes as many modes as possible to lower its energy. Finally, for $V<0$ or $U_{s}<0$, $F_{{\rm toy}}$ is not bounded from below and therefore unstable.

To make use of this toy model, we must tune the interaction such that $U(\theta,\phi)\to0$ unless $\theta=\pi/2$ or $\phi=0$. This removes the quartet term $F_{\rm nt}$ and we are left  with a model similar to $F_{{\rm toy}}$ with the difference that $V_{p}(\theta)$ is not constant.

We now tune the interaction such that $V_{p}(\theta)$ is very big\cite{verybig} everywhere except at  some angle $\theta_{\rm opt}$,
 where it has a  narrow minimum. For such a pair interaction, all arrangements of modes which involve angles other than $2\theta_{\rm opt}$ are  excluded. Within this constraint, we are left with our toy model with a constant pair interaction $V=V_{p}(\theta_{\rm opt})$, but the number of modes is restricted to $1\leq N\leq 3$, since no more than three modes can have equal mutual angles between them. It is clear that in the region $-3U_{s}/2<V_{p}(\theta_{\rm opt})<U_{s}$, the ground state is a helical spin crystal with three modes, i.e., the state ``$\bigtriangleup$'' or sc (in the case $2\theta_{\rm opt}=\pi/2$).

A different class of exact ground states is obtained, if
the interaction is tuned such that
\begin{enumerate}
\item $V_{p}(\theta)$ is very big\cite{verybig} in the region $0<2\theta<2\theta_c$ and
\item  $V_{p}(\theta)=V$ (constant) in the region  $2\theta_c\leq2\theta\leq\pi/2$
\end{enumerate}
for some critical angle $2\theta_c$. In this way, modes whose wavevectors are too close are excluded. That is, it enforces a ``hard sphere'' constraint  $|\uv k_i\pm\uv k_j|\geq 2(1-\cos{2\theta_c})$ on the wavevectors.
The interaction between modes which satisfy this constraint is constant and  reduces to the toy model. Therefore, if $V<U_s$ the ground state will include as many modes as geometrically possible by the ``hard sphere'' constraint. In the case $2\theta_c=\arccos 1/3$, we obtain fcc, which has  wavevectors at the vertices of a cube. In the case $2\theta_c=\pi/3$, we obtain bcc whose wavevectors are the vertices of a cuboctahedron. This can be seen as follows.
A real-space bcc lattice corresponds to a fcc reciprocal lattice and the wavevectors $\pm k_j$ of the bcc spin crystals are the  12 nearest neighbors of the origin in the fcc reciprocal lattice. Because fcc is the cubic close packing of spheres, this represents the only arrangement of twelve vectors $|\bv k_j|=Q$  which satisfies the constraint $|\uv k_i-\uv k_j|\geq1$. (The hexagonal close packing is not acceptable because it does not consist of pairs $\pm\bv k_j$.) 
 In principle, this construction can be used to create models whose ground state contains an arbitrarily high number of modes.

We can now use the insight from these constructed models to understand certain features of the phase diagram of Figs.~\ref{pdiag0} and \ref{pdiag-2}.
In Fig. \ref{Vpfig}, we plot $V_{p}(\theta)$ as obtained from the expansion Eq.~\eref{Ulm} at different places in the phase diagram of Fig.~\ref{pdiag0}. In the bcc1 region of the phase diagram (curve C),  $V_{p}(\theta)$
is similar to the one constructed above: close to constant and small for  $\pi/3\leq 2\theta\leq\pi/2$, big for $2\theta<\pi/3$. At the phase boundary between single-spiral and bcc1 (curve B), $V_{p}$ is still bigger than $U_{s}$, indicating that the quartet term $F_{\rm nt}$ is essential to stabilize bcc1 close to the phase boundary. The phase boundaries between single-spiral and three-mode states (sc or $\bigtriangleup$) are exactly determined by the crossing of the minimum of $V_{p}(\theta)$ with $U_{s}$, as illustrated by curves D and E.

\section{Structure of helical spin crystals}\label{structure}

\subsection{Symmetry properties and topology of helical spin crystals.}\label{symmetry}

Time-reversal symmetry  ($\mathcal{T}$)  reverses the magnetization direction $\bv M\to-\bv M$ and may be implemented in terms of the $\psi$-variables as $\psi_j\to-\psi_j$.

Under spatial translation by a vector $\bv a$, the $\psi$-variables transform as $\psi_j'=\psi_j\exp(i\bv k_j\bv a)$, i.e., they experience a phase change.

If the set of vectors $\bv k_j$ is linearly independent, every phase-change of modes simply amounts to a global translation. This is the case in helical spin crystals with up to three non-planar modes, e.g. in sc and $\bigtriangleup$.  These states are periodic with Bravais vectors $\bv a_1,\bv a_2,\bv a_3$, which are determined by $\bv a_i\cdot\bv k_j=2\pi\delta_{ij}$. It follows that  time-reversal $\mathcal{T}$ is  equivalent to a translation by $(\bv a_1+\bv a_2+\bv a_3)/2$ for these states, which leads to a 2-fold symmetry inside  the unit cell reminiscent of antiferromagnetism.

In contrast, helical spin crystals with more than three modes like bcc, fcc and $\square$ depend essentially on the relative phases between the helical modes. For these states,  $\mathcal{T}$ is {\em not}  equivalent to any translation, since it is not possible that $\exp(i\bv k_j\bv a)=-1$ for all $\bv k_j$.  As a consequence, these states are doubly degenerate in addition to the translational degeneracy.

To obtain an understanding of the real-space picture of helical spin crystals, it is useful to study their symmetries under rotations (reflections change the chirality and therefore never appear in the symmetry group).   For example, if the structure $\bv M(\bv r)$ has a $n-$fold rotation axis with direction $\uv u$, then $\bv M\parallel\uv u$ along this axis.
Therefore,  $\bv M_\perp$, which is the projection  to the plane orthogonal to $\uv u$, has a node at the axis.
 The  winding number\cite{mermin79} (vorticity) of the node is restricted by symmetry 
 to the values $1,1\pm n,1\pm 2n$, etc.  
 The resulting pattern in the vicinity of the rotation axis, $\bv M$ pointing along $\uv u$ in the center and $\bv M_\perp$ winding around it, resembles that of a skyrmion or meron. Such patterns are currently being discussed  in the context of helimagnets.\cite{bogdanov05,fischer06} Here, we observe that they naturally appear resulting from rotational symmetries.
The simplest cases are winding numbers of $+1$ (for any $n$-fold axis)  or $-1$ (only for $n=2$).

Apart from proper rotations, the point group may contain anti-rotations, i.e.,  symmetry operations which are composed of a rotation followed by $\mathcal{T}$. $n$-fold anti-rotation axes are only possible for even numbers $n$, since they imply  $n/2$-fold rotation symmetry. In the special case $n=2$, it merely follows that $\bv M\perp\uv u$ along the axis, where $\uv u$ is the  axis direction. Higher anti-rotation symmetries with $n=4,6,\ldots$ imply $\bv M=0$ along the axis, i.e., they create a line-node in the magnetization. In the vicinity of this line-node, $\bv M$ is approximately orthogonal to $\uv u$ (as can be seen by expanding $\bv M(\bv r)$ to linear order around a point on the axis). The winding number of $\bv M_\perp$ around the node, is restricted  by symmetry to the values $1\pm n/2,1\pm3n/2,1\pm5n/2$, etc.     
 The simplest case for a 4-fold anti-rotation axis is a winding number of $-1$ (anti-vortex line) and  the simplest case for $n=6$ is a winding number of $-2$.

We have thus demonstrated the emergence of topological objects like merons and antivortices, which are not expected to be stable in the present context of a vectorial order parameter in three dimensions but which are stabilized by symmetry. Thus, rotation axes are meron lines and 4-fold anti-rotation axes are anti-vortex node-lines.

\subsubsection{Symmetry and real-space picture of bcc1}\label{bcc1}

\begin{figure}
\includegraphics[scale=0.6]{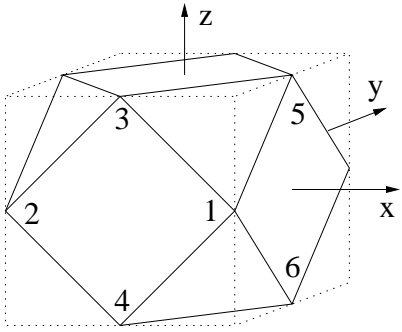}
\caption{Geometry of the wavevectors $\bv k_1,\ldots,\bv k_6$, which constitute the bcc spin crystal states. The vectors $\pm \bv k_j$ form the vertices of a cuboctahedron.
\label{6modes}}
\end{figure}

The bcc helical spin crystals consist of six helical modes with wavevectors $\uv k_1=[1\bar10]$,  $\uv k_2=[\bar1\bar10]$,  $\uv k_3=[0\bar11]$,  $\uv k_4=[0\bar1\bar1]$,  $\uv k_5=[101]$ and  $\uv k_6=[10\bar1]$, as shown in Fig.~\ref{6modes}, and has the periodicity of a body centered cubic lattice.
 We chose the convention [see Eq.~\eref{psis}]
\be
\uv \epsilon_j''=\frac{\uv z\times\uv k_j}{|\uv z\times\uv k_j|},\label{conv}
\ee
for $j=1,\ldots,6$ with $\uv z=[001]$, and we consider negative chirality [i.e., $\bv M\cdot(\nabla\times\bv M)<0$], such that $\uv \epsilon_j'=\uv k_j\times\uv \epsilon_j''$.
The geometry of the wavevectors allows for three quartet terms in the free energy produced by $T_x=\psi_1^*\psi_2\psi_5\psi_6$,
$T_y=\psi_1^*\psi_2^*\psi_3\psi_4$ and
$T_z=-\psi_3\psi_4^*\psi_5^*\psi_6$.  The sign in the definition of $T_z$ has been introduced for convenience.

The transformation properties of the $\psi$-variables and the three quartic terms  under rotations are shown in Table \ref{trans}. From this, it can be deduced that the (rotation-invariant) quartet contribution to the free energy is
\be
F_{\rm nt}=2\lambda_{\rm nt}\, \Re(T_x+T_y+T_z).
\ee
The value of the parameter $\lambda_{\rm nt}$ is not determined by symmetry. Direct calculation yields
\be
\lambda_{\rm nt}=\frac74U[\frac\pi3,2\arccos(\frac13)]-\frac12U(\frac\pi4,\pi).
\ee 

There are thus two bcc states, depending on the sign of $\lambda_{\rm nt}$. In bcc1 (bcc2), which corresponds to $\lambda_{\rm nt}>0$ ($<0$), the phases of the $\psi$'s are such that $T_x$, $T_y$, $T_z$ are all negative (positive). Three out of the six phases are arbitrary due to global translation symmetry. This means that the magnetic pattern of bcc1 and bcc2 is uniquely determined up to translational and time-reversal degeneracy.

\begin{table}
\begin{ruledtabular}
\begin{tabular}{c|cccccc|ccc}
 & $\psi_1'$ &  $\psi_2'$ &  $\psi_3'$ &  $\psi_4'$ &  $\psi_5'$ &  $\psi_6'$ & $T_x'$ & $T_y'$ & $T_z'$ \\
\hline
$R_z$ & $\psi_2$ & $\psi_1^*$  & $\psi_6^*$ & $\psi_5^*$ & $\psi_3$ & $\psi_4$ & $T_y$ & $T_x^*$ & $T_z^*$ \\
$R_x$ & $i\psi_5$ & $-i\psi_6^*$  & $-\psi_4^*$ & $\psi_3$ & $i\psi_2^*$ & $i\psi_1$ &  $T_x^*$ & $T_z$ & $T_y^*$
\end{tabular}
\caption{Transformation properties of the $\psi$-variables and three quartic terms (defined in Section \ref{bcc1}) of the bcc spin crystals under rotations. $R_z$ and $R_x$, respectively, are $\pi/2$ rotations around the $z$- and $x$-axis. These two rotations generate the cubic point group $O$ and therefore, the behavior under any rotation which maps the 12 wavevectors onto each other may be obtained by combining these two operations.
\label{trans}}
\end{ruledtabular}
\end{table}

The solution for $\lambda_{\rm nt}>0$, bcc1, turns out to be the bcc structure with the highest point group symmetry.
By selecting  the coordinate origin conveniently, we obtain  
 $-\psi_1=\psi_2=-i\psi_3=i\psi_4=i\psi_5=-i\psi_6=SM_0$ for bcc1, where $S=\pm1$ is the time-reversal symmetry label and $M_0>0$ is the amplitude. From Table \ref{trans}, we deduce that $\bv M(\bv r)$ changes sign under a $\pi/2$ rotation about the  $x$, $y$ or $z$ axis. That is, the magnetic point group is $O(T)$ (international notation $\underline{4}32$) with 4-fold anti-rotation axes  at $\langle 100\rangle$, 3-fold rotation axes at $\langle111\rangle$ and 2-fold anti-rotation axes at $\langle110\rangle$.

\begin{figure}
\includegraphics[scale=0.6]{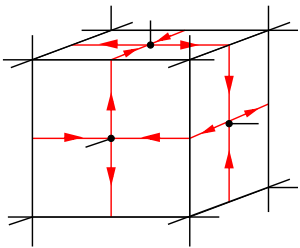}
\caption{(Color online). Symmetry of the bcc1 state. The figure shows a cubic unit cell of bcc1.  Black lines are anti-vortex lines with 4-fold anti-rotation symmetry and vanishing magnetization. The red (dark gray) lines are 2-fold rotation axes and the arrows indicate the direction of $\bv M$. The structure has 3-fold rotation symmetry about all cubic space diagonals.
\label{real-space}}
\end{figure}

The real-space representation of the bcc1 state is
\be
\bv M(\bv r)=SM_0\left(\begin{array}{c}\sqrt2\,s_x(c_y-c_z)-2\,s_ys_z\\ \sqrt2\,s_y(c_z-c_x)-2\,s_zs_x\\ \sqrt2\,s_z(c_x-c_y)-2\,s_xs_y\end{array}\right),
\ee
 where $s_x=\sin(Qx/\sqrt{2})$,  $c_x=\cos(Qx/\sqrt{2})$, etc.
 The resulting pattern was shown in Fig.~2 of our earlier paper.\cite{binz06} In Fig.~\ref{real-space}, we show the symmetry axes. As discussed above, the magnetization must vanish along the 4-fold anti-rotation axes, which are anti-vortices with winding number $-1$. The  $x$, $y$, $z$ axes, and their translations according to the bcc periodicity, form two interpenetrating cubic latices of such line-nodes. The  cubic space diagonals are 3-fold and the red arrowed lines of Fig.~\ref{real-space} are 2-fold  rotation axes.
 In the vicinity of these lines, the magnetization field is skyrmion-like (i.e., $\bv M_\perp$ has  a winding number of $+1$).

The fact that the bcc1 state breaks $\mathcal{T}$ in a non-trivial way and cannot be restored by any translation is manifest in the occurrence of the $\mathcal{T}$-breaking order parameter $\langle M_xM_yM_z\rangle=SM_0^3/2\neq0$, which is a magnetic octupole. This curious property may lead to distinctive anomalous effects, e.g. in the magnetotransport.\cite{binz06b} 
Octupolar magnetic ordering has recently been discussed in different contexts.\cite{octupole}

\subsubsection{Symmetry and real-space picture of sc}

The simple cubic (sc) helical spin crystal consists of three modes with  $\uv k_1=[100]$,  $\uv k_2=[010]$ and  $\uv k_3=[001]$. It forms a periodic structure with a cubic unit cell and the lattice constant is $\lambda=2\pi/Q$.
 The convention for $ \epsilon_j''$ is given by  Eq.~\eref{conv}, where  the unit vector $\uv z$ is replaced by $[111]$.

\begin{table}
\begin{ruledtabular}
\begin{tabular}{c|ccccc}
& $R_x$ & $R_y$ & $R_z$ & $R_{[111]}$ & $R_{[1\bar 1 0]}$ \\
\hline
 $\psi_1'$ & $i\psi_1$ & $-i\psi_3$ & $\psi_2^*$ & $\psi_3$ & $-\psi_2^*$ \\
 $\psi_2'$ & $\psi_3^*$ & $i\psi_2$ & $-i\psi_1$ & $\psi_1$ & $-\psi_1^*$ \\
 $\psi_3'$ & $-i\psi_2$ & $\psi_1^*$ & $i\psi_3$ & $\psi_2$ & $-\psi_3^*$
\end{tabular}
\caption{Transformation for the $\psi$-variables of sc under rotations. $R_z$ and $R_x$ are $\pi/2$-rotations, $R_{[111]}$ is a $2\pi/3$-rotation and $R_{[1\bar 1 0]}$ a $\pi$-rotation around the indicated axis.
\label{transsc}}
\end{ruledtabular}
\end{table}

 The transformation properties of the $\psi$-variables under rotations are given in Table \ref{transsc}. By choosing the center of coordinates corresponding to $\psi_1=\psi_2=\psi_3=iM_0$, we obtain from Table \ref{transsc} that the  point group symmetry is $D_3(D_3)$ (international notation 32). That is, the chosen origin has a 3-fold rotation axis along $[111]$ and three two-fold axes along $[1\bar10]$, $[10\bar1]$ and $[01\bar1]$. Obviously, $\bv M$ must vanish at a point of such high symmetry. Hence there is a point node at the origin.

Symmetry operations consisting of a rotation followed by an appropriate translation yield similar point nodes at $\frac 12\frac {3}4\frac {1}4$ (with 3-fold axis along $[\bar111]$), $\frac 14\frac {1}2\frac 34$ (3-fold axis $[1\bar11]$) and   $\frac {3}4\frac {1}4\frac {1}2$ (3-fold axis $[11\bar1]$). Finally, each of these nodes is doubled inside one unit cell
 because a translation by $(\frac \lambda2,\frac \lambda2,\frac \lambda2)$ amounts to $\bv M\to-\bv M$.
The 2- and 3-fold rotation axes form a  complex array of skyrmion-like lines, all with winding numbers of $+1$.
The real-space representation is
\be
\bv M(\bv r)=SM_0\left(\begin{array}{c}
\tilde c_y-\tilde s_z\\
\tilde c_z-\tilde s_x\\
\tilde c_x-\tilde s_y
\end{array}\right),
\ee
where  $\tilde s_x=\sin[Q(x+\lambda/8)]$,   $\tilde c_x=\cos[Q (x+\lambda/8)]$, etc.

\subsection{Higher harmonics Fourier modes}\label{higher-harmonics}

As briefly mentioned in Section \ref{continuous}, magnetic ordering in wavevectors $\pm\bv k_j$ generally induces higher harmonics in the magnetic structure. In the presence of magnetic order $\bv m_{\bv k_j}\neq0$,  the Landau free energy for the modes $\bv m_{\bv q}$, which do not belong to the set  $\bv m_{\bv k_j}$, is (to quartic order)
\be
\Delta F=\sum_{\bv q} \tilde r_\psi(\bv q) |\bv m_{\bv q}|^2-\bv h_{\psi}(\bv q)\cdot \bv m^*_{\bv q}-\bv h^*_{\psi}(\bv q)\cdot \bv m_{\bv q}, \label{induce}
\ee
with $\tilde r_\psi(\bv q)=r(q)+O\left(|\psi_j|^2\right)$ and
\be
\bv h_{\psi}(\bv q)=-4
 \!\!{\sum_{\bv q_1,\bv q_2,\bv q_3}}^{\!\!\!\!\!\!\prime}\,\,U(\bv
q_1,\bv q_2,\bv q_3)\left(\bv m_{\bv q_1}\cdot\bv m_{\bv
q_2}\right)\,\bv m_{\bv q_3},
\label{hpsi}
\ee
where the sum  is restricted to $\bv q_1,\bv q_2,\bv q_3\in \{\pm\bv k_{j}\}$ such that $\bv q_1+\bv q_2+\bv q_3=\bv q$. The origin of the exchange field $\bv h_{\psi}$ is  the  coupling term  Eq.~\eref{F4}. In the following, we assume that $\tilde r(\bv q)>0$.  Obviously, Eq.~\eref{induce} then leads  to induced modes
\be
\bv m_{\bv q}=\frac{\bv h_{\psi,\bv q}}{\tilde r(\bv q)}\label{harmonics}
\ee
at momenta $\bv q=\pm\bv k_{j_1}\pm\bv k_{j_2}\pm\bv k_{j_3}$.\footnote{In non-magnetic crystals, the coupling term is cubic and therefore  higher harmonics are generated at all sums of {\em two} momenta $\bv k_{j_1}\pm\bv k_{j_2}$. In magnetic structures, 
 higher-harmonics wavevectors are restricted to  sums of {\em three} ordering momenta.} These modes modify the detailed magnetic structure, but they do not change its symmetry, since the field $\bv h_\psi$ respects all the symmetries of the spin crystal.

We now briefly discuss the consequences for the three helical magnetic structures under consideration.

 A single spin-density wave involving wavevectors $\pm\bv k$ might create  higher harmonics at $\pm3\bv k$ via  Eq.~\eref{harmonics}. However, in the case of spin spirals, $\bv m_{\bv k}^2=0$ and therefore $\bv h_{\psi,3\bv k}=0$.
 Thus, there are  {\em no} higher harmonics created by a single spin spiral.

The sc spin structure with principal ordering wave-vectors along $\langle001\rangle$ with $|\bv k_j|=Q$
 generates higher harmonics along $\langle111\rangle$  (with $|\bv q|=\sqrt{3}Q$) and
 along $\langle012\rangle$   (with $|\bv q|=\sqrt{5}Q$). Note that throughout the current and last sections, all crystal directions refer to the magnetic crystals.
The orientation of a magnetic crystal with respect to the atomic crystal depends on the  anisotropy term $F_{\rm a}$, which will be considered in  Section \ref{crystalanisotropy}.

In contrast to the former cases, bcc structures couple linearly to the $\bv q=0$ mode (i.e., the uniform magnetization), since some triples of ordering vectors  add to zero. This coupling will be further investigated in Section \ref{H}. Here, we only notice that for the  bcc1 and bcc2 states,  $\bv h_{\psi,\bv q=0}=0$. This results can be understood in terms of the symmetry of these states. In the case of bcc1, the point group symmetry is too high to support a non-zero axial vector $\bv h_{\psi,\bv q=0}$. Therefore, bcc1 and bcc2 do not create a spontaneous net magnetization.
The next set of wavevectors which can be reached by adding three ordering vectors are along $\langle001\rangle$ (with $|\bv q|=\sqrt{2}Q$).  However  for bcc1 and bcc2,  direct calculation shows  $\bv h_{\psi}=0$ for these modes. As before, this can be understood in terms of symmetry. Higher harmonics along  $\langle001\rangle$  would  have the structure of a  sc spin crystal. We have seen in Section \ref{symmetry}, that the point group symmetry of sc is lower than that of bcc1. Therefore, bcc1 can not create such an exchange field $\bv h_{\psi}$. 
 We conclude that bcc1 creates {\em no} secondary Bragg peaks at  $(0,0,\sqrt{2}Q)$, etc. The same is true for bcc2.
The shortest wavevectors which are created by bcc1 or bcc2 as higher harmonics are along $\langle 112\rangle$ (with $|\bv q|=\sqrt{3}Q$). Others are at  $\langle110\rangle$ ($|\bv q|=2Q$),   $\langle013\rangle$ ($|\bv q|=\sqrt{5}Q$),  $\langle111\rangle$ ($|\bv q|=\sqrt{6}Q$) and  $\langle123\rangle$ ($|\bv q|=\sqrt{7}Q$).

\section{Response to crystal anisotropy, magnetic field and disorder.}\label{response}

\subsection{Effect of crystal anisotropy}

So far, our  free energy has
been completely rotation invariant. In the magnetically ordered states,  full rotation symmetry
is spontaneously broken, but any global rotation of the spin
structure leaves the energy invariant.
 This degeneracy is lifted by an additional anisotropy term  $F_{\rm a}$, which couples the  magnetic crystal to  the underlying atomic lattice. The crystal anisotropy energy is small and may be treated as a perturbation which merely selects the directional orientation, but does not otherwise affect the magnetic state.

\subsubsection{Single-spiral state}\label{spiral-anisotropy}

In the case of a single-spiral state, crystal anisotropy is a function $F_{\rm a}(\uv k)$, where $\uv k$ is the spiral direction.
 The function  $F_{\rm a}(\uv k)$ may depend on various parameters, it should be symmetric under the
 point group of the (atomic) crystal lattice and satisfy $F_{\rm a}(\uv k)=F_{\rm a}(-\uv k)$. For concreteness, we assume the cubic point group $T$, relevant for the $B20$ crystal structure. We further assume that $F_{\rm a}(\uv k)$ is a slowly varying function, since a singular or rapidly oscillating function in reciprocal space would translate into a (non-local) interaction between magnetic moments and the atomic crystal. Such a function $F_{\rm a}(\uv k)$ generally has its minimum at either $\left\langle100\right\rangle$ or $\langle111\rangle$, which can be shown in two different ways.

The first argument is based on combining symmetry with Morse's theory of critical points.\cite{morse34}  Morse theory implies that
\be
\mbox{maxima}-\mbox{saddles}+\mbox{minima}=2 \label{morse}
\ee
for a function on the unit sphere. Symmetry requires that $F_{\rm a}(\uv k)$ has stationary points (points with vanishing first derivative, i.e., maxima, minima or saddles) at $\left\langle100\right\rangle$ (6 directions),  $\left\langle111\right\rangle$ (8 directions) and  $\left\langle110\right\rangle$ (12 directions).
If  $F_{\rm a}(\uv k)$ is slowly varying, we  suspect that these are the only stationary points, since adding more maxima, minima and saddles means that the function is more rapidly oscillating. Under this hypothesis, it follows from Eq.~\eref{morse}, that the  $\left\langle110\right\rangle$ directions are saddle points and that the extrema are at  $\left\langle111\right\rangle$ and  $\left\langle100\right\rangle$. For   $\left\langle110\right\rangle$ to be minima,  $F_{\rm a}(\uv k)$ needs to have additional stationary points (e.g. saddles) at non-symmetric, parameter-dependend locations. We conclude that an anisotropy  which favors  $\left\langle110\right\rangle$ would need to be more rapidly oscillating than required by symmetry.

The second argument is based on an expansion of  $F_{\rm a}(\uv k)$ in powers of the directions cosines $\hat k_x, \hat k_y,\hat k_z$:
\be
 F_{\rm a}(\uv k)=\alpha\, (\hat k_x^4+\hat k_y^4+\hat k_z^4) + \alpha' \, \hat k_x^2\hat k_y^2\hat k_z^2+\ldots,\label{Fa}
\ee
where we retained the first two terms allowed by cubic symmetry. Because of the smallness of the wave-vector sphere $Q$, one  typically expects $|\alpha'|\ll |\alpha|$ and subsequent terms even smaller. It is easily checked that for most values of the parameters $\alpha,\alpha'$, Eq.~\ref{Fa} has its global minima at either  $\left\langle100\right\rangle$ (for $\alpha<\min\{0,\alpha'/18\}$) or  $\left\langle100\right\rangle$ (for $a>\max\{2\alpha'/9,\alpha'/18\}$). Only in the narrow parameter regime $0<\alpha<2\alpha'/9$, the minima are indeed at $\left\langle110\right\rangle$.\footnote{In this regime, all 26 high-symmetry points are extrema of $F_{\rm a}$ and 24 saddle points are located at $\left\langle\sqrt{2\alpha}\,\sqrt{2\alpha}\,\sqrt{\alpha'-4\alpha}\right\rangle$, in agreement with Eq.~\eref{morse}.} We conclude that crystal anisotropy  which favors  $\left\langle110\right\rangle$ may only appear in a narrow regime between two phases which favor  $\left\langle111\right\rangle$ and   $\left\langle100\right\rangle$, respectively.

Accordingly,   $\left\langle111\right\rangle$ or   $\left\langle100\right\rangle$ are the selected spiral directions in all cubic helimagnets known so far.\cite{ishikawa76,beille83,lebech89}
The preferred direction in MnSi at low pressure is $\left\langle 111\right\rangle$ and in  Fe$_x$Co$_{1-x}$Si, it is  $\left\langle 100\right\rangle$.  In FeGe, there is a phase transition between these two directions,\cite{lebech89} but no  intermediate phase with  $\left\langle110\right\rangle$ spiral orientation has been reported. However,  the neutron scattering data\cite{pfleiderer04}  in the partially ordered phase of MnSi clearly show a maximum signal along the $\left\langle110\right\rangle$ crystal directions. While it is initially tempting to  interpret the partially ordered state of MnSi as a  single-spiral state that has lost it's orientational long-range order by some mechanism,  one would still expect a maximal scattering intensity in the energetically preferred lattice direction.
 Theories of the partially ordered state in terms of disordered helical spin-density waves\cite{tewari05,grigoriev05} thus depend on a crystal anisotropy that prefers spiral directions along $\left\langle 110\right\rangle$. As we have shown, this seems very unlikely  .

\subsubsection{Helical spin crystals}\label{crystalanisotropy}

 For multi-mode spin crystals, 
 $F_{\rm a}$  is no longer determined by a single direction $\uv k$, so the arguments of Section \ref{spiral-anisotropy} do not apply.  Rather, the anisotropy energy depends on  three Euler angles, which rotate the full three-dimensional magnetic structure relatively to the atomic crystal. In other words, $F_{\rm a}$  is a function of the rotation group $SO(3)$. Relative to some standard orientation $\bv k_j$ of the mode directions,  the  leading-order anisotropy term is
\be
F_{\rm a}(R)=a\sum_{j}g({R\uv k_j})\,|\psi_j|^2
\label{crysFa}
\ee
where 
 $R$ is a rotation operator and   $g(\uv k)=\hat k_x^4+\hat k_y^4+\hat k_z^4$. As before, we have assumed a cubic point group symmetry.

In the case $a>0$, the modes of the bcc  spin crystals get locked to the  $\langle110\rangle$ directions. The  orientation of sc is four times degenerate if $a>0$. The four minima of $F_{\rm a}$ are obtained from the standard orientation along $\langle100\rangle$ through a $\pi/3$-rotation around any of the four space diagonals, such that the three spiral modes point along  $\langle122\rangle$.

In the opposite case ($a<0$), sc is oriented along $\langle100\rangle$. This time, it is the bcc spin crystals that get rotated by  $\pi/3$  around any  $\langle111\rangle$ axis to reach one of four stable orientations.  Under such  $\pi/3$-rotations, three of the six modes remain along $\langle110\rangle$  and three  move to $\langle114\rangle$. Each individual  $\langle114\rangle$ direction appears only in one of the four solutions but each  $\langle110\rangle$ direction appears in two of four solutions.

If there is more than one degenerate
 orientation, the sample typically breaks up into domains such that  full cubic symmetry is restored in the neutron scattering signal.  Table \ref{orienttable} lists the directions of magnetic Bragg peaks for the 
 different cases.
 Out of the three prominent phases in our phase diagram,  the bcc1 spin crystal is the only one  that can explain the neuron-scattering peaks along $\langle110\rangle$ in the 'partial order' phase of MnSi. It does so most naturally for the case $a>0$, which is the known sign of the anisotropy in MnSi at low pressure.

In order to compare the energy scale of $F_{\rm a}$ (i.e., the locking energy) for the different magnetic states, we note the following.
 At the phase boundary between two equal-amplitude spin-crystal phases (one phase with amplitudes $|\psi_1|=\ldots=|\psi_N|$ and the second phase with $|\tilde\psi_1|=\ldots=|\tilde\psi_{\tilde N}|$)
the amplitudes of the two neighboring phases are related by
\be
N|\psi_j|^2=\tilde N|\tilde\psi_{j'}|^2.
\ee
In the vicinity of the phase boundary, the anisotropy term is therefore proportional to the mode-average  $1/N\sum_j g(R\uv k_j)$. Using this result, we find  that the effective anisotropy energy is smaller for the bcc spin crystals than for the single-spiral state by a factor of 4 - 4.5.\footnote{Using $\max\{F_{\rm a}\}-\min\{F_{\rm a}\}$ to estimate the locking energy scale leads to a reduction factor of 4.5. Expanding $F_{\rm a}$ around its minimum for $a>0$ leads to a  factor of 4.}
For the sc state, the locking energy is  anisotropic. Certain rotations are equally costly in energy as for the single-spiral case while some small rotations about the minimum for $a>0$ are softer than for the single-spiral  by a factor of 4.5.

\begin{table}
\begin{ruledtabular}
\begin{tabular}{c|cc|cc|cc}
 & spiral & No.  & bcc & No.  & sc & No.  \\
\hline
 $a>0$ &$\langle111\rangle$ & 4 & $\langle110\rangle$ & 1 &$\langle122\rangle$ & 4\\
 $a<0$ &$\langle100\rangle$ & 3 & $\langle110\rangle$,$\langle114\rangle$\footnote{When averaged over domains, Bragg peaks along $\langle110\rangle$ are twice as intense as peaks along $\langle114\rangle$.} & 4 &$\langle100\rangle$ & 1
\end{tabular}
\caption{Crystal directions of magnetic Bragg peaks and number  of degenerate orientations for three magnetic structures, sc, bcc and single-spiral, with crystal anisotropy given by Eq.~\eref{crysFa}.
\label{orienttable}}
\end{ruledtabular}
\end{table}

\subsection{Effect of magnetic field}\label{H}

A uniform external magnetic field $\bv H$ couples to the $\bv q=0$ mode of the
magnetization, $\bv m=\langle\bv M\rangle$, via Zeeman coupling.
The uniform magnetization, in turn, couples to the helical modes $\psi_j$ through
\be
\begin{split}
F=&\left.F\right|_{\bv m=0}+r_0\,\bv m^2+U(0,0,0)\,\bv m^4-\bv h_{\psi}(0)\,\bv m\\
&+2\sum_j\left(U(\bv k_j,-\bv k_j,0)\,\bv m^2+U(\bv k_j,0,0)\,\bv m^2_{\perp,j}\right)|\psi_j|^2,\label{Fm}
\end{split}
\ee
where  $\bv m_{\perp,j}=\bv m-(\bv m\cdot\uv k_j)\uv k_j$ and we have used Eqs.~\eref{F4},\eref{hpsi}.  Plumer and Walker\cite{plumer81} argued that $U(0,0,0)\approx U(\bv k_j,-\bv k_j,0)\approx U_{s}$, which we will use in the following for simplicity.

\subsubsection{Response of the single-spiral state}

The behavior of the single-mode spiral state under a
magnetic field has been studied both experimentally\cite{ishikawa76,lebech89,ishimoto95, thessieu97,uchida06} and theoretically.\cite{kataoka81,plumer81}
 It is characterized by a strongly anisotropic susceptibility,
induced by the last term in Eq.~\eref{Fm}. For a fixed spiral direction $\bv k$, the  susceptibilities parallel and orthogonal to the spiral direction are given by
\be
\begin{split}
\chi_\parallel\,\approx&\,\Delta^{-1}\\
\chi_\perp\approx&\left(\Delta+2\frac{U'}{U_{s}}|r(Q)|\right)^{-1},
\end{split}
\ee
where $\Delta=2[r_0-r(Q)]=2JQ^2$ and $U'=U(\bv k,0,0)$. Well below the critical ordering temperature, $\Delta\ll 2|r(Q)|$ and therefore $\chi_\parallel\gg \chi_\perp$. This strong anisotropy leads to a spin reorientation transition at $H=H_{\rm sr}$, where the spiral axis gets oriented along the field direction. The value of $H_{\rm sr}$ depends on the anisotropy [Eq.~\eref{Fa}] and the field direction. For $\alpha>0$ and $\bv H\parallel \langle 100\rangle$, Plumer and Walker obtained
\be
H_{\rm sr}^2=\frac{4\alpha}{\chi_\parallel-\chi_\perp}\gtrsim 4 \alpha \Delta,\label{Hsr}
\ee
where we have used  $\chi_\parallel\gg \chi_\perp>0$. Once the spiral is oriented, the susceptibility is large (equal to $\chi_\parallel$).

The spiral amplitude decreases as a function of the external field and vanishes at
\be
 H_{\rm c}=|\psi_0|\Delta,\label{Hc}
\ee
where $|\psi_0|^2=|r(Q)|/(2U_{s})$. Above $H_{\rm c}$, the magnetization is uniform.

\subsubsection{Response of the bcc1 spin crystal}\label{bcc-and-field}

 In the bcc spin crystal
states, the linear response is isotropic, because
their symmetry group does not allow for an anisotropic
susceptibility tensor. As a consequence, there is no orientation of the bcc state towards the magnetic field at the level of linear response (i.e., from energies up to order $\bv H^2$). However, there is a sub-leading contribution to the energy $\propto\langle M_xM_yM_z\rangle H_xH_yH_z$. This contribution splits the degeneracy between the $S=1$ and $S=-1$ states and it may lead to a reorientation of the bcc crystal towards the field.

In terms of the six $\psi$-variables of bcc, the exchange field $\bv h_\psi(0)$, which enters Eq.~\eref{Fm}, amounts to
 \be
\bv h_\psi(0)=-\mu \Re{\left[(5-i\sqrt{2})\bv{\tilde h}_\psi\right]},
\ee
where $\mu=U(\bv k_{j_1},\bv k_{j_2},\bv k_{j_3})$ for $\bv k_{j_1}+\bv k_{j_2}+\bv k_{j_3}=0$ and
\be
\begin{split}
\bv{\tilde h}_\psi= & \,\psi_1\psi_3^*\psi_6^*\left(\begin{array}{c}1\\1\\1\end{array}\right)
+\psi_1^*\psi_4\psi_5\left(\begin{array}{c}-1\\-1\\1\end{array}\right) \\
& -\psi_2\psi_4^*\psi_6\left(\begin{array}{c}-1\\1\\-1\end{array}\right)
-\psi_2^*\psi_3\psi_5^*\left(\begin{array}{c}1\\-1\\-1\end{array}\right).
\end{split}
\ee

The (isotropic)  inverse spin susceptibility (see Appendix) in the bcc1 state is composed of three contributions
\be
\chi^{-1}_{\rm bcc1}=\chi^{-1}_{\rm bare}+\chi_{\rm phase}^{-1}+\chi_{\rm amp}^{-1}.\label{chibcc1}
\ee
The first term
\be
\chi^{-1}_{\rm bare}=2\left(r_0+\frac{U_s+\frac23 U'}{U_{\rm bcc1}}|r(Q)|\right),
\ee
where $U_{\rm bcc1}=1/6[U_{s}+V_{p}(\pi/4)+4V_{p}(\pi/6)-\lambda_{\rm nt}]$,  can be derived in analogy to the single-spiral case. In fact, $\chi_{\rm bare}$ is a ``mixture'' of $\chi_\parallel$ and $\chi_\perp$, determined geometrically by the angles between the mode directions $\bv k_j$ and the magnetic field. It follows that $\chi_{bare}\ll \chi_\parallel$, provided $U_{\rm bcc1}\sim U_{s}$ (the two couplings are equal at the phase boundary between single-spiral and bcc1).
The remaining terms in Eq.~\eref{chibcc1}
\be
\begin{split}
\chi_{\rm phase}^{-1}&=-\,\frac{\mu^2\,|r(Q)|}{3\,U_{\rm bcc1}\,\lambda_{\rm nt}}\\
\chi_{\rm amp}^{-1}&=-\,\frac{25\,\mu^2\,|r(Q)|}{12\,U_{\rm bcc1}\,[U_{s}-V_{p}(\pi/4)+\lambda_{\rm nt}]},
\end{split}\label{chi-phase-amp}
\ee
stem from the response of the bcc magnetic structure to the field. That is, they originate from the adjustments of relative phases and amplitudes, respectively, of the helical modes as a result of the term $-\bv h_\psi(0)\cdot\bv m$ in  Eq.~\eref{Fm}. The effect of $\chi_{\rm phase}$ and $\chi_{\rm amp}$, which are necessarily negative, is to increase the susceptibility of the bcc1 state.

The change in
the relative amplitudes and phases of the six interfering spirals as
a function of the magnetic field may  be calculated  (see Appendix). For example, the
linear response of the amplitudes of bcc1 is
\be
\left(\begin{array}{c}
\delta|\psi_1|\\ \delta|\psi_2|\\ \delta|\psi_3|\\ \delta|\psi_4|\\ \delta|\psi_5|\\ \delta|\psi_6|\\
\end{array}\right)=\frac{5\,\mu\, S}{4 [U_{s}-V_{p}(\pi/4)+\lambda_{\rm nt}]}\left(\begin{array}{r}
-m_z\\ m_z\\ -m_x\\ m_x\\ m_y\\ -m_y
\end{array}\right).
\ee
This response should be observable by neutron scattering, if it is
possible to prepare the sample in a single-domain state (i.e.
without mixture of the two time-reversal partners). For example, a field in $\uv z$ direction affects $|\psi_1|$ and $|\psi_2|$, the amplitudes of the modes propagating orthogonally to $\uv z$ (Fig. \ref{6modes}), which get enhanced and suppressed by the magnetic field, respectively. 

The expected effects of external magnetic field on the resistivity of bcc spin crystal are presented elsewhere.\cite{binz06b}

\section{Effect of impurities: a possible route to 'partial order'.}\label{disorder}

While the helical spin crystal states are expected to show Bragg
spots at particular wavevectors, a variety of effects such as
thermal or quantum fluctuations or disorder can destroy the long
range order while preserving the helical spin crystal structure at
shorter scales. Here, we investigate in more detail the effect of
non-magnetic disorder on helical spin crystal structures.

 Although
the experimentally studied helimagnets are very clean from the
electrical resistivity point of view, the helical magnetic
structures are sensitive to disorder at a much longer length scale.
In addition, the low energy scales required to distort them means
that one needs to consider disorder effects. An observation that can
immediately be made is that for the physically relevant case of
non-magnetic disorder ($V_{dis}(\bv r)$), the single-spiral state
and the spin crystal states respond very differently. By symmetry,
the coupling of disorder to the magnetic structure is given by
$F_{dis}=\langle V_{dis}(\bv r) |\bv M(\bv r)|^2\rangle$. Hence,
single-spiral states which are unique in having a spatially uniform
magnitude of magnetization ($|\bv M(\bv r)|={\rm constant}$) are
unaffected by this coupling; in contrast the spin-crystal states
necessarily have a modulated magnitude\cite{binz06} and hence are
affected by non-magnetic disorder. Therefore the neutron scattering
signal of the spin-crystal state is expected to have more diffuse
scattering than the single mode state. This is consistent with the
experimental observation that the high pressure phase has diffuse
scattering peaked about $\langle110\rangle$ while the low pressure
phase has sharper spots, consistent with identifying the two as
spin-crystal and single-spiral states respectively.

 The effect of disorder on the
spin-crystal state is closely related to the problem
of the ordering of an XY model in the presence of a random external field.
The phase rotation symmetry of the XY model captures the
translational invariance of the spin-crystal in the clean state.
Disorder destroys this invariance and behaves like a random field
applied to the XY system. Using the insights from the study of that
problem in three dimensions,\cite{giamarchi95}
 one expects that for weak
disorder a Bragg glass will result, where although true long range
order is destroyed, power law divergent peaks at the Bragg
wavevectors remain, and the elastic constants remain finite. For
stronger disorder one expects this algebraic phase to also be
destroyed, and recover a short range correlated phase without
elasticity. Nevertheless, for the case of the bcc1 and bcc2
crystals, due to time reversal symmetry ($\mathcal{T}$) breaking in
these states, the disordered states also spontaneously break time
reversal symmetry, and hence a phase transition is expected on
cooling despite the absence of long range order. It is difficult to
predict which of these two scenarios (Bragg glass or only
$\mathcal{T}$ breaking) is more appropriate for MnSi. In the latter
case one may estimate the spreading of the Bragg spots due to
disorder by considering the energetic cost to deform the
spin-crystal state in different ways.

Ignoring elastic contributions,
 there are two
distinct types of deformations - ones that involve a change in the
magnitude of the ordering wavevectors $\delta q_\parallel$
 and others that do not change
the wavevector magnitude but rotate the structure from its preferred
orientation: $\delta q_\perp$. The second is expected to be low in energy because
rotations of the structure are locked by the crystal anisotropy
term, which is weak.
 From Eq.~\eref{Fa}, we obtain the energy cost to shift the ordering vector  by $\delta q_\perp$ along the sphere $|\bv q|=Q$
\be
\delta F_\perp=\frac{4\alpha}{3\kappa}\left(\frac{\delta  q_\perp}Q\right)^2,\label{Fperp}
\ee
where $\kappa=1$ for the single-spiral. For multi-mode spin crystals, the energy cost of rotation is reduced, as explained in Section \ref{crystalanisotropy}. Thus, $\kappa\approx4$ for the bcc spin crystals.
In contrast, deformations that change the magnitude of the ordering
wavevectors, must contend with the DM
interaction scale, and hence pay a higher energy penalty
\be
\delta F_\parallel=\frac12\Delta\cdot|\psi_0|^2\cdot\left(\frac{\delta  q_\parallel}Q\right)^2.\label{Fpar}
\ee
Assuming
the disorder couples to these deformations equally, we can estimate
the ratio of their amplitudes in the limit of weak deformations, by equating Eqs.~\eref{Fperp} and \eref{Fpar}. It follows
\be
\left(\frac{\delta  q_\perp}{\delta  q_\parallel}\right)^2=\frac{3\kappa\Delta|\psi_0|^2}{8\alpha}.
\ee
Using Eqs.~\eref{Hsr} and \eref{Hc}, we can relate this ratio to the experimentally known ratio between the critical magnetic fields for, respectively, reorienting and  polarizing the single-spiral state
\be
\frac{\delta  q_\perp}{\delta  q_\parallel}
\gtrsim\sqrt{\frac{3\kappa}2}\frac{H_{\rm c}}{H_{\rm sr}}.
\ee

We can now apply these results to the case of MnSi and test the
hypothesis that the 'partial order' state is in fact a
disordered bcc spin crystal. Setting in $\kappa=4$ and the
experimentally measured\cite{thessieu97} critical fields for MnSi,
$H_{c}=0.6{\rm Tesla}$ and $H_{c1}=0.1{\rm Tesla}$, one obtains
$\delta q_{\perp}/\delta q_\parallel\gtrsim 15$. Neutron scattering
experiments do indeed find that the transverse broadening is larger
than the longitudinal broadening, but since the latter is resolution
limited, this only gives us an lower bound that is consistent with
the estimate above: $[\delta q_{\perp}/ \delta
q_{\parallel}]_{\rm{expt}}> 2.3$. Nevertheless, the trend that the
width of the spot is greater along the equal magnitude sphere than
transverse to it is clearly seen in the experimental data.

Thus, weak non-magnetic disorder of the atomic crystal is expected to destroy magnetic long range order in multi-mode helical spin crystal states and lead to a neutron scattering signal compatible with the observations in the 'partial order' phase of MnSi. 
 However in the case of bcc spin crystals, time-reversal symmetry breaking is expected to persist even in the presence of disorder. The scenario of interpreting 'partial order' in MnSi as a bcc1 state disordered by impurities thus predicts  quasi-static local magnetic moments 
  and implies  a finite temperature phase transition on cooling into this phase. 
 
\section{Conclusion}

We have analyzed the magnetic properties of non-centrosymmetric weak ferromagnets subject to DM spin-orbit coupling. This problem falls into the general class of systems where the low energy excitations live on a surface in reciprocal space rather than on discrete points.
The addition of DM interactions to a ferromagnetic state produces a
large degeneracy of magnetic states characterized by arbitrary
superpositions of spin helices of a fixed helicity and fixed
wavevector magnitude. This enormous degeneracy is broken by
interactions between modes, and the single-spiral state is realized
for slowly varying interactions, by virtue of its unique property of
having a spatially uniform magnitude of magnetization. For more
general interactions, multi-mode helical spin crystal states are
obtained. We show that for the model interactions considered, the
phase diagram is largely determined just by considering the
interactions between pairs of modes. The phase that is eventually
realized may be readily deduced from the range of angles in which
this interaction drops below a critical value. In particular, the
bcc structure is stabilized by virtue of the fact that its
reciprocal lattice, fcc, is a close packed structure. These results
may also be relevant in other physical situations where
crystallization occurs, such as the Larkin-Ovchinnikov-Fulde-Ferrel
 instability in spin-imbalanced superconductors, which may
potentially be realized in solid state systems,\cite{CeCoIn5} cold
atomic gases\cite{zwierlein} and dense nuclear matter.\cite{Rajagopal}

Helical spin crystals typically give rise to complicated real space
magnetic structures which we discussed in this paper. In particular,
topological textures like merons and anti-vortices can be seen about
special axes in particular realizations, although these are not
expected to be stable given the order parameter and spatial
dimensionality of the system. We show here that such topological
structures exist as a consequence of symmetry, which also dictates
the absence of certain higher Bragg reflections, which a naive
analysis would predict.

The response of helical spin crystals to crystalline anisotropy and
applied magnetic field are considered with a special emphasis on the
bcc structures which are contrasted against the response of the
single helix state. An unusual transfer of spectral intensity in the
presence of an applied magnetic field, which is strongly dependent
on the direction of applied field is noted for the bcc structures.
This is a consequence of broken time reversal symmetry in the
absence of a net magnetization (which is symmetry forbidden). The
unusual magnetotransport in such a state, a linear in field
magnetoresistance and quadratic Hall effect, has been discussed
briefly in Ref.~\cite{binz06} and was elaborated upon in Ref.~\cite{binz06b}.

 Helical spin crystals exhibit Bragg peaks at specific
wave-vectors, and hence are not directly consistent with the
experimental observation of 'partial order'. The point of view taken
in our earlier work\cite{binz06} is that the short distance and
short time properties are captured by the appropriate helical spin
crystal structure. Studying the properties of helical spin crystals
with long range order is a theoretically well defined task with
direct consequences for a proximate disordered phase with similar
correlations up to some intermediate scale. The mechanism
that leads to the destruction of long range helical spin crystal
order is uncear; in Ref.~\cite{binz06}, this was assumed to be the coupling
to non-magnetic disorder. Then, as elaborated in this paper,
beginning with a bcc helical spin crystal a neutron scattering
signature consistent with that of 'partial-order' may be obtained.
However, within the simplest version of this scenario, one also
expects a finite temperature phase transition where time reversal
symmetry breaking develops, and static magnetic order which may be
seen in nuclear magnetic resonance or muon spin rotation
experiments. Other mechanism for the destruction of long range order
of the bcc spin crystal state, such as thermal or quantum
fluctuations may also be considered, but are left for future work.

\acknowledgments We would like to thank L. Balents, P. Fazekas, I. Fischer, D.
Huse, J. Moore, N. Nagaosa, M.P. Ong, C. Pfleiderer, D. Podolsky, A. Rosch, T.
Senthil, H. Tsunetsugu and C. Varma for useful and stimulating discussions and V.
Aji for an earlier collaboration on related topics. This work was
supported by the Swiss National Science Foundation, the A.P. Sloan
Foundation and Grant No. DE-AC02-05CH11231.

\appendix*

\section{Linear response}\label{appendix}

Let the free energy $F$ depend on real internal variables $\bv x=(x_1,\ldots,x_n)$ (related to various order parameters)  and on $\bv m=(m_1,m_2,m_3)$, which couples linearly to the external field $\bv H$. The internal variables are chosen such that for $\bv m=0$, the minimum energy shall be at $\bv x=0$.
 Expanding $F$ to second order in $\bv x$ and $\bv m$ yields
\be
F=\frac12\bv x^TA\,\bv x+\bv m^T B\, \bv x+\frac12\bv m^T C\, \bv m,\label{Flr}
\ee
where $A$, $B$ and $C$ are matrices, $A=A^T$, $C=C^T$. All eigenvalues of $A$ are positive. Eq.~\eref{Flr}
 can be written as
\be
F=\frac12(\bv x-\bv x_m)^TA\,(\bv x-\bv x_m)+\frac12\bv m^T\chi^{-1}\bv m.
\ee
where  $\bv x_m=-A^{-1}B^T\bv m$ and  $\chi^{-1}=C-BA^{-1}B^T$. It follows that under the influence of an external field $\bv H$, the equilibrium  internal variables get shifted to $\bv x=\bv x_m$ and the linear response is given by  $\bv m=\chi \bv H$. Thus, there are two contributions to the inverse suszeptibility: $\chi_{\rm bare}^{-1}=C$ and $\chi_{x}^{-1}=-BA^{-1}B^T$. The latter comes from the internal response of the $x$-variables to the magnetic field.  

These general results are applied in Section \ref{bcc-and-field}, where the internal variables $x_1,\ldots,x_9$ are the deviations from their equilibrium value at $\bv m=0$ of six amplitudes $|\psi_j|$ and  three phases (holding the other three phases fixed). In this case, $\chi_x$ leads to both the phase and amplitude related terms in Eq.~\eref{chi-phase-amp}.


\begin{thebibliography}{0}
\bibitem{verybig} Without having a rigorous proof, we suspect that in the context of Section \ref{exact}, ``very big'' just means ``bigger that $U_{s}$''. This bound has proven sufficient in all cases that we have tested.
\end{thebibliography}
\end{document}